\documentclass[12pt,preprint]{aastex}

\begin{document}

\title{Supernova-Triggered Molecular Cloud Core Collapse and 
the Rayleigh-Taylor Fingers that Polluted the Solar Nebula}

\author{Alan P.~Boss and  Sandra A. Keiser}
\affil{Department of Terrestrial Magnetism, Carnegie Institution,
5241 Broad Branch Road, NW, Washington, DC 20015-1305}
\authoremail{boss@dtm.ciw.edu, keiser@dtm.ciw.edu}

\begin{abstract}

 A supernova is a likely source of short-lived radioisotopes (SLRIs) 
that were present during the formation of the earliest solar
system solids. A suitably thin and dense supernova shock wave may be 
capable of triggering the self-gravitational collapse of a molecular 
cloud core while simultaneously injecting SLRIs. Axisymmetric hydrodynamics
models have shown that this injection occurs through a number of
Rayleigh-Taylor (RT) rings. Here we use the FLASH adaptive mesh
refinement (AMR) hydrodynamics code to calculate the first fully 
three dimensional (3D) models of the triggering and injection process. 
The axisymmetric RT rings become RT fingers in 3D. While $\sim 100$ 
RT fingers appear early in the 3D models, only a few RT fingers are 
likely to impact the densest portion of the collapsing cloud core.
These few RT fingers must then be the source of any SLRI spatial 
heterogeneity in the solar nebula inferred from isotopic analyses of 
chondritic meteorites. The models show that SLRI injection efficiencies 
from a supernova several pc away fall at the lower end of the range
estimated for matching SLRI abundances, perhaps putting them more into
agreement with recent reassessments of the level of $^{60}$Fe present 
in the solar nebula.

\end{abstract}


\section{Introduction}

 Chondritic meteorites contain daughter products of the decay of
short-lived radioisotopes (SLRIs), such as $^{26}$Al (Lee et al. 
1976; Amelin et al. 2002) and $^{60}$Fe (Tachibana \& Huss 2003),
present during the formation of  
the earliest solids in the solar system. Recent evidence has 
raised doubts about the initial amount of $^{60}$Fe present in the
solar nebula. Moynier et al. (2011) presented evidence for an
initial ratio of $^{60}$Fe/$^{56}$Fe less than $3 \times 10^{-9}$,
roughly 100 times smaller than previously published ratios (e.g.,
Mishra et al. 2010), while Tang \& Dauphas (2012) found an initial
ratio of $1 \times 10^{-8}$. The initial level of $^{60}$Fe 
is particularly significant, as its efficient production requires 
stellar  nucleosynthesis (Tachibana et al. 2006) 
in either a massive star supernova or an AGB star (Huss et al. 
2009). If the initial level of $^{60}$Fe is low enough, cosmic rays
may be sufficient to create this SLRI (Moynier et al. 2011). The
interstellar medium (ISM) appears to have 
$^{60}$Fe/$^{56}$Fe = $1 \times 10^{-7}$
(Tang \& Dauphas 2012), implying that lower ratios could be 
inherited from the ISM after a suitable decay interval 
(e.g., Gounelle et al. 2009; Tang \& Dauphas 2012). 
While the initial $^{60}$Fe/$^{56}$Fe ratio is
uncertain, other recent studies have shown that this ratio appears 
to be $\sim 2 \times 10^{-7}$ for certain chondrules found in
unequilibrated ordinary chondrites (Telus et al. 2012). Such 
ratios may require a stellar nucleosynthetic source for this SLRI. 

 Boss et al. (2010) used the FLASH AMR code to study
presolar cloud core triggering and injection 
processes associated with SLRI production in either a core collapse 
supernova or an AGB star. They showed that shock waves from a 
supernova or an AGB star could simultaneously trigger the collapse of 
a dense molecular cloud core and inject shock wave material 
into the resulting protostar. Boss \& Keiser (2010, hereafter BK10) 
found, however, that the injection efficiency depended sensitively 
on the assumed shock width and density. Supernova shock waves 
appeared to be thin enough to inject the desired amount of shock wave 
material. Planetary nebula shock waves, derived from AGB star winds, 
however, were too thick to achieve the desired injection 
efficiencies. BK10 thus concluded that a supernova
was the likely trigger for solar system formation, a conclusion
that has been supported by subsequent theoretical (Gritschneder 
et al. 2012), observational (e.g., Diehl et al. 2010; Phillips
\& Marquez-Lugo 2010), and cosmochemical studies (e.g., Young et al. 
2011; Krot et al. 2012). Other scenarios are reviewed by Adams 
(2010) and Boss (2012).

\section{Numerical Methods and Initial Conditions}

 Here we extend the BK10 2D AMR models  
to fully 3D AMR calculations using the 3D Cartesian coordinate 
($x, y, z$) version of FLASH2.5 in much the same manner as our
previous calculations with the axisymmetric ($R$, $Z$) version.
The grid was 0.2 pc long in $y$ (the direction along
which the shock wave travels initially) and 0.13 pc wide in $x$
and $z$. The target cloud is a Bonnor-Ebert sphere with a central
density of $1.24 \times 10^{-18}$ g cm$^{-3}$, a radius of 0.058 pc,
and a mass of 3.8 $M_\odot$, initially centered on the grid
at $x = z = 0$ and $y =$ 0.13 pc. The cloud is composed of molecular
hydrogen gas with a mean molecular weight $\mu = 2.3$. 
The initial number of blocks in $x$ and $z$ was 6 and in $y$ was 9, 
with each block consisting of $8^3$ grid points. 
The number of levels of grid refinement was initially 4, but was 
increased to as high as 7 levels (when possible) to better resolve
the protostellar core, leading to effective resolutions as
high as $3072^3$ in regions with large gradients in the density
or color (shock front) fields. A typical simulation 
ran for three months on the dedicated flash cluster at DTM.

 The initial shock parameters (Table 1) overlap with those of BK10, 
where the standard shock number density was $n_s = 10^4$ 
cm$^{-3}$ and shock width was $w_s = 10^{16}$ cm. Model 40-400-0.1 
has $v_s =$ 40 km/sec, shock number density $n_s = 4 \times 10^6$ cm$^{-3}$,
and shock width $w_s = 10^{15}$ cm. The post-shock density $n_s$ for an isothermal shock in a gas of density $n_m$ is $n_s/n_m = (v_s/c_m)^2$, 
where $c_m$ is the pre-shock sound speed. For our 40 km/sec models, 
with $c_m = 0.2$ km/sec and $n_m = 10^2$ cm$^{-3}$, $n_s = 4 \times 
10^6$ cm$^{-3}$, the same density as in model 40-400-0.1. For comparison, 
W44 is a Type II supernova remnant (SNR) with $v_s = 20-30$ km/sec, 
width $< 10^{16}$ cm, and radius $\sim 11$ pc, expanding into gas with 
$n \sim 10^2$ cm$^{-3}$ (Reach et al. 2005).

 As in BK10, we included compressional heating and radiative cooling, 
based on the results of Neufeld \& Kaufman (1993) for
cooling caused by rotational and vibrational transitions of optically 
thin, warm molecular gas composed of H$_2$O, CO, and H$_2$, leading to 
a radiative cooling rate of $\Lambda \approx 9 \times 10^{19} (T/100) 
\rho^2$ erg cm$^{-3}$ s$^{-1}$, where $T$ is the gas temperature in K 
and $\rho$ is the gas density in g cm$^{-3}$.  

\section{Results}

 The overall evolution of the 3D AMR models proceeds  in a manner 
very similar to that of the 2D AMR models of Boss et al. (2008, 2010) 
and BK10: the shock wave strikes
the target cloud core, compressing the facing edge of the cloud,
as the shock wave propagates unimpeded around the sides of the cloud,
forming a parabolic shock front. Figure 1 shows a cross-section
through model 40-200-0.1 after 0.02 Myr of evolution, when 
the cloud core has been compressed by a factor of $\sim$ 30
by a shock initially propagating downwards (toward $y = 0$).
Model 40-200-0.1 has $v_s =$ 40 km/sec, $\rho_s$ = $7.2 \times 
10^{-18}$ g cm$^{-3}$ (200 times the standard density), and 
$w_s =$ 0.0003 pc (0.1 times the standard width). Figure 1 shows 
that the color field, representing the shock wave material, has 
been injected into the shock-compressed outer cloud layers. The shock  
corrugates the surface of the compressed cloud core with a series 
of indentations indicative of a Rayleigh-Taylor (RT) instability.
Meanwhile, Kelvin-Helmholtz (KH) instabilities caused by velocity 
shear form in the relatively unimpeded portions of the shock 
front at large impact parameter compared to the center of the
target cloud. The shock front gas and dust is
entrained and injected into the compressed region by the RT
instability, while the KH rolls result in ablation and loss
of target cloud mass in the downstream flow. 

 Figure 1 shows that in 3D, the initially high symmetrical 
configuration remains fairly axisymmetrical. The number of
RT fingers evident in the cross section at this time
is $\sim 16$, the same number of fingers (assuming axisymmetry) as
seen in the corresponding BK10 2D AMR model,
and with a very similar spacing and distribution across the 
shock/cloud interface. The main difference is that the 2D RT 
structures are rings, not fingers. While we have not performed
a convergence study (e.g., Pittard et al. 2009; Yirak et al. 2010),
the fact that the 2D model has nearly four times the spatial resolution
in each direction as the 3D model, yet yields the same number of
R-T features, implies a reasonable degree of numerical convergence.
Figure 2 shows a cross-section in the direction perpendicular to that 
of the shock wave propagation (i.e., in the $x$ - $z$ plane), which 
clearly exhibits the formation of numerous ($\sim$ 100) distinct RT 
fingers in the shock wave matter. 

Figure 3 shows model 40-200-0.1 after 0.063 Myr
of evolution, when the cloud core has been compressed to a
maximum density over $10^5$ times higher than the initial maximum
density: the cloud core is dynamically collapsing with a radius of 
order 100 AU. As in 2D, by this time the color field  has been injected 
throughout the collapsing region, though with varying color field 
density. The mass of the collapsing protostar is roughly $1 M_\odot$, 
with a maximum density at this time of $\sim 2 \times 10^{-13}$
g cm$^{-3}$, high enough to enter the regime
of protostellar collapse when the collapsing cloud core becomes
optically thick in the infrared, and our assumption of optically thin,
radiative cooling begins to fail. 

Figure 4 shows the status of the RT fingers at the same time
(0.063 Myr) as Figure 3, displayed again in the $x - z$ plane,
so that the number of RT fingers can be counted. Figure 4
is plotted for $y = 0.045$ pc = $1.4 \times 10^{17}$ cm, at a 
height just above the protostellar object seen in Figure 3.
Figure 4 shows that the 100-odd RT fingers evident in Figure 2
have been reduced to about 10, largely due to the ablation of target
cloud mass into the downstream flow. Figure 4 shows that only 
a limited number of RT fingers will inject SLRIs.
In fact, given that the radius of the protostar at this time is 
$\sim$ 100 AU $\sim 10^{15}$ cm, Figure 4 shows that it is
possible that a single RT finger, the one at the center, will
successfully inject shock wave material into the protostar.
The remaining RT fingers will likely continue downstream past
the collapsing protostar. This reduction in the number of
RT fingers responsible for injection also appears
to be due in part to merging of the fingers, as Figure
4 shows that the RT fingers are less distinct than in
Figure 2. A similar effect can be seen in the high resolution 2D 
models of Vanhala \& Boss (2002): the number of distinct RT rings 
decreases from $\sim$ 12 at 0.022 Myr to $\sim$ 4 at 0.13 Myr.

With the exception of model 40-800-0.1, where the shock-compressed
shell was shredded and did not undergo collapse, all the other models
listed in Table 1 evolved in a manner similar to that of
model 40-200-0.1 and displayed in the four figures.
The result for model 40-800-0.1 was to be expected, as the
same model failed to undergo sustained collapse when calculated
in 2D by BK10. It is reassuring, however, to obtain the same result 
when re-calculated in 3D, as the implication is that 2D models 
can be used to survey a larger parameter space than can be explored 
with 3D models.

\section{Injection Efficiency and Dilution Factors}

For comparison to isotopic analyses of primitive meteorites,
we must estimate injection efficiencies and dilution factors.
We assume that the SLRIs are carried by dust grains that are small 
enough to move along with the gas, as calculated by the AMR code.
The injection efficiency $f_i$ is defined to be the fraction of 
the incident shock wave material that is injected into the collapsing 
cloud core. The factor $\beta$ is the ratio of shock front mass
originating in the SN to the mass swept up in the intervening ISM.
The dilution factor $D$ is then defined as the ratio of the 
amount of mass derived from the supernova to the amount of 
mass derived from the target cloud. The portion of the 
shock front in model 40-200-0.1 incident on the cloud
contains 0.3 $M_\odot$ of gas and dust, implying
$D \approx 0.3 \beta f_i$, for a final system mass of 1 $M_\odot$.
This assumes that the injected SLRIs are uniformly 
distributed in the collapsing cloud core, as well as in the
resulting protostar and protoplanetary disk. 

Estimates of the injection efficiency $f_i$ are made difficult
by the fact that by the time that the calculations are halted
because of the rising optical depth, the protostar is still
far from having formed a well-defined young stellar object
and protoplanetary disk. Hence the exact amount of shock front
matter that is incorporated into the protostar is uncertain, 
as it will be a combination of the matter already injected into 
the collapsing region and that accreted at a later time. Given this
uncertainty, the injection efficiencies for the 3D models appear
to be similar to those of the corresponding BK10 2D models:
the color field densities in the collapsing region of the 40 km/sec 
shock model 200-0.1 (BK10's Figure 4) are very similar ($\sim 0.03$ in 
dimensionless units, compared to an initial color density of 1) 
to that of model 40-200-0.1 at the time shown in Figures 3 and 4.

For model 40-200-0.1, assuming that a region around the collapsing 
protostar seen in Figure 3 with a radius of $\sim 10^{16}$ cm
is eventually accreted by the protostar, the injection
efficiency $f_i \sim 0.03$, close to the BK10 estimate
of $f_i \sim 0.02$. With this new estimate for $f_i$, we find 
$D \sim 0.01 \beta$. In order for a supernova shock to slow down to a
speed of 40 km/sec, however, a considerable amount of ISM matter 
must be snowplowed by the shock front, reducing $D$ by a factor of 
$\beta \sim 0.01$ to $D \sim 10^{-4}$, for a shock that travels
$\sim 5$ pc from a 20 $M_\oplus$ SN (e.g., Ouellette  et al. 2007) 
through an ISM with $n \sim 10^2$ cm$^{-3}$. This estimate falls at the
low end of the range of dilution factors inferred for typical SLRIs 
from SN, namely $\sim 10^{-4}$ to $\sim 3 \times 10^{-3}$ 
(Takigawa et al. 2008; Gaidos et al. 2009). 

Several other factors, however, could result in increased values 
of $D$, such as enhanced mixing associated with sub-grid turbulence 
(Pittard et al. 2009), preferential addition of the SLRIs to the 
disk rather than the protostar (BK10), and enhanced
injection of the SLRIs through their presence in dust grains
large enough to punch through the shock-cloud interface more
effectively than gaseous RT fingers (BK10). While investigation
of the intermediate scenario will require calculations of rotating
target clouds, so that protoplanetary disks can form, the latter
scenario can be evaluated now. The core-collapse 
supernova 1987A, e.g., appears to have produced a dust mass 
of about 0.4 to 0.7 $M_\odot$ (Matsuura et al. 2011), and
these dust grains presumably carry the freshly synthesized SLRIs.

BK10 estimated that dust grains larger than $\sim 30 \mu$m
would be needed in order to increase $D$ values. However, dust grains in
SNRs are thought to be smaller than $\sim 30 \mu$m. Nozawa et al. (2010) 
found an average radius of dust grains less than 0.01 $\mu$m
for the Cas A SNR of a Type IIb SN.
Andersen et al. (2011) fit the spectral energy distributions
of 14 SNRs with several different populations of dust grains, where
the largest grains needed were smaller than 0.1 $\mu$m. In their models 
of dust processing in C-type shocks, appropriate for SNRs, Guillet 
et al. (2011) considered dust grains all smaller than 0.03 $\mu$m. 
Hence dust grains in SNRs do not appear to be large enough to result
in significantly enhanced injection efficiencies. Another constraint
comes from presolar dust grains, derived from a variety of stellar outflows,
such as SN (e.g., Amari et al. 1994) and AGB stars (e.g., Bernatowicz 
et al. 2006), some of which can be 6 $\mu$m in size. However, only 
about 1\% of these grains are larger than 1 $\mu$m. We conclude that SLRIs 
carried by large dust grains are insufficient to raise the injection
efficiencies significantly.

 We are left, however, with another means of reconciling 
the dilution factors estimated on the basis of these 3D models 
($D \sim 10^{-4}$) with those inferred from a combination of SN
nucleosynthesis abundance calculations and laboratory analysis of 
primitive meteorites, which have been as high as $\sim$ 30 times 
larger (Takigawa et al. 2008; Gaidos et al. 2009). 
The production amounts of SLRI such as 
$^{26}$Al and $^{60}$Fe in core-collapse supernovae appear to 
be uncertain by factors of five or more (Tur et al. 2010). In
addition, the fact that recent estimates of the initial $^{60}$Fe
abundance in the solar nebula have fallen considerably, in
some cases by factors of 100 or more (Moynier et al. 2011),
suggests that it may well be possible to
accommodate the supernova triggering and injection scenario
for the origin of the solar nebula's SLRIs. $^{60}$Fe seems 
to be the key SLRI, as $^{26}$Al is relatively abundant in 
the interstellar medium (e.g., Diehl et al. 2010) and is
expected to be concentrated in the outflows of Wolf-Rayet
stars that are the predecessors to many Type II SN
(Tatischeff, Duprat, \& de S\'er\'eville 2010).
AGB stars, however, still appear to be ruled out, on the
basis of the even lower injection efficiencies resulting
from the much thicker width of planetary nebula outflows
compared to SNRs (BK10).

\section{Conclusions}

While supernova triggering and injection is 
a possible explanation of the evidence for SLRIs in meteorites,
it remains to be seen if a combination of target cloud and shock 
front parameters can be found that will produce the correct SLRI 
injection efficiencies and also be consistent with observations of SNRs.
Assuming the suitability of our shock front parameters,
the 3D models show that a relatively small number of RT
fingers are likely to have been involved in the injection
of the SNR SLRIs into the solar nebula, which has significant
consequences for the processes of mixing and transport that occurred
after injection into the nebula (e.g., Boss 2011) and for the
resulting levels of isotopic homogeneity and heterogeneity
(e.g., Bouvier \& Wadhwa 2010; Schiller et al. 2010; Larsen et al. 
2011; Makide et al. 2011; Wang et al. 2011; Boss 2012).
Furthermore, our estimates of injection efficiencies
and dilution factors will become important discriminators for
judging the likelihood of the supernova triggering and injection
scenario once a clearer picture emerges of the initial abundances
of the SLRIs (principally $^{60}$Fe: Moynier et al. 2011;
Telus et al. 2012) that were present during the formation of the
various components of the most primitive meteorites.

\acknowledgements

 We thank the referees for their help, Larry 
Nittler, Scott Messenger, and Jeff Cuzzi for discussions, and Michael 
Acierno and Ben Pandit for assistance with the flash cluster at DTM.
This research was supported in part by NASA Origins of Solar Systems 
grant NNX09AF62G and contributed in part to NASA Astrobiology Institute
grant NNA09DA81A. The software used in this work was in part
developed by the DOE-supported ASC/Alliances Center for 
Astrophysical Thermonuclear Flashes at the University of Chicago.

\clearpage
\begin{deluxetable}{lccccccc}
\tablecaption{Initial parameters explored for the 3D AMR models, showing
the outcomes for models with varied initial shock speeds ($v_s$, in
km/sec), shock density ($\rho_s$, in units of the standard density 
of $3.6 \times 10^{-20}$ g cm$^{-3}$), and shock width ($w_s$, in 
units of the standard width of 0.003 pc). The outcomes are listed
as C for sustained collapse and NC for no sustained collapse.
\label{tbl-1}}
\tablewidth{0pt}
\tablehead{\colhead{shock density $\times$} 
& \colhead{} 
& \colhead{1} 
& \colhead{200}
& \colhead{400}
& \colhead{800}
& \colhead{1000}
& \colhead{1600} }
\startdata

shock width $\times$  1  & $v_s =$ 20  & C & -  & -  & -  & - & -  \\

shock width $\times$ 0.1 & $v_s =$ 20  & - & C  & C  & C  & C & C  \\

shock width $\times$  1  & $v_s =$ 40  & C & -  & -  & -   & - & - \\

shock width $\times$ 0.1 & $v_s =$ 40  & - & C  & C  & NC  & - & - \\

\enddata
\end{deluxetable}

\clearpage

\begin{figure}
\vspace{-1.0in}
\includegraphics[scale=.80]{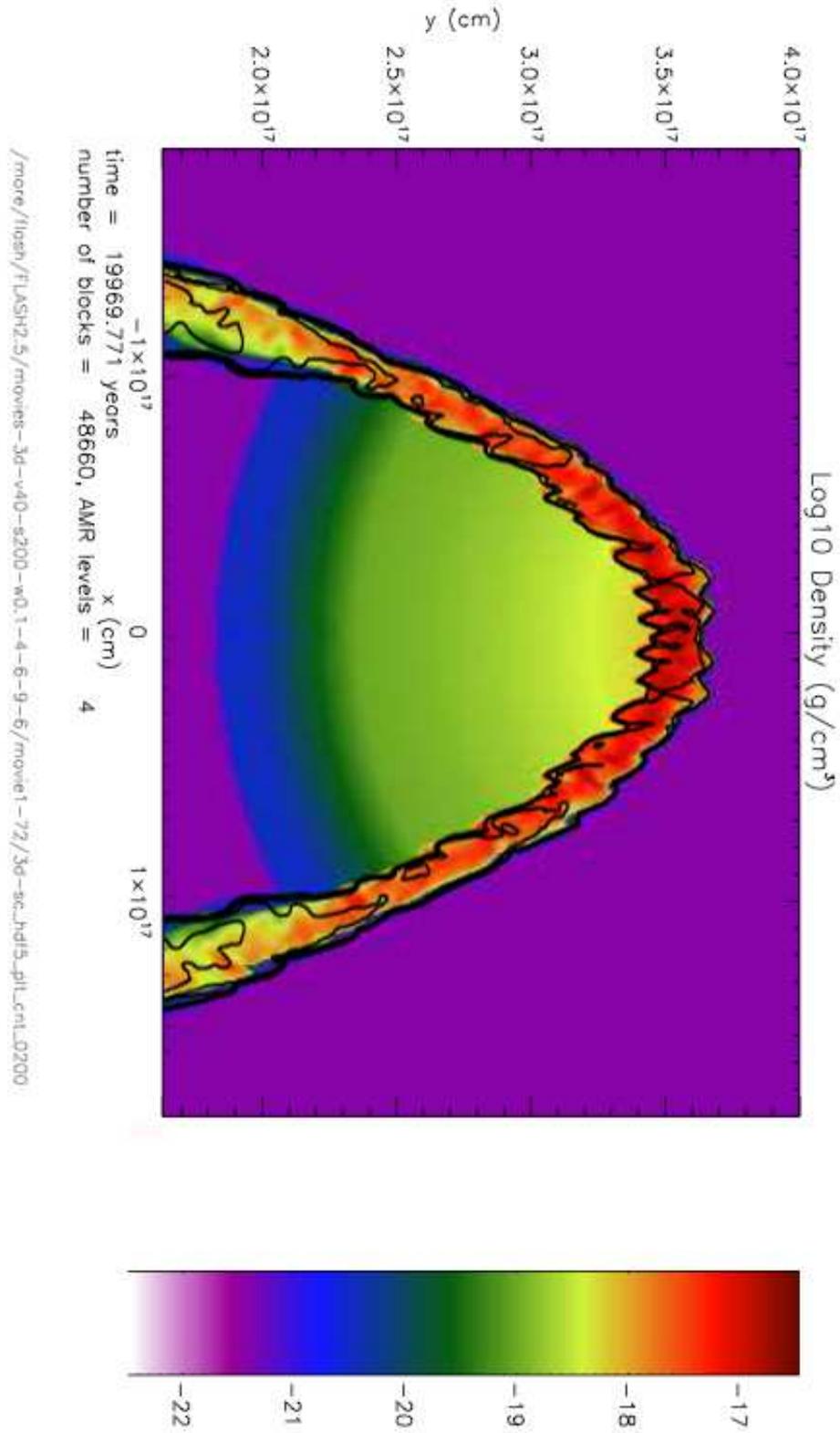}
\vspace{+0.25in}
\caption{Log10 of the density distribution for model 40-200-0.1 
after $2.00 \times 10^{4}$ yrs of evolution, plotted in the $z = 0$
plane. Contours show the color field plotted at 0.01, 0.025, 0.05, 
0.075, and 0.1. The $x$ axis is horizontal and the $y$ axis is vertical.
The downward propagating shock wave has compressed the target
cloud core and is injecting shock front material through
multiple Rayleigh-Taylor (RT) fingers.}
\end{figure}

\begin{figure}
\vspace{-1.0in}
\includegraphics[scale=.80]{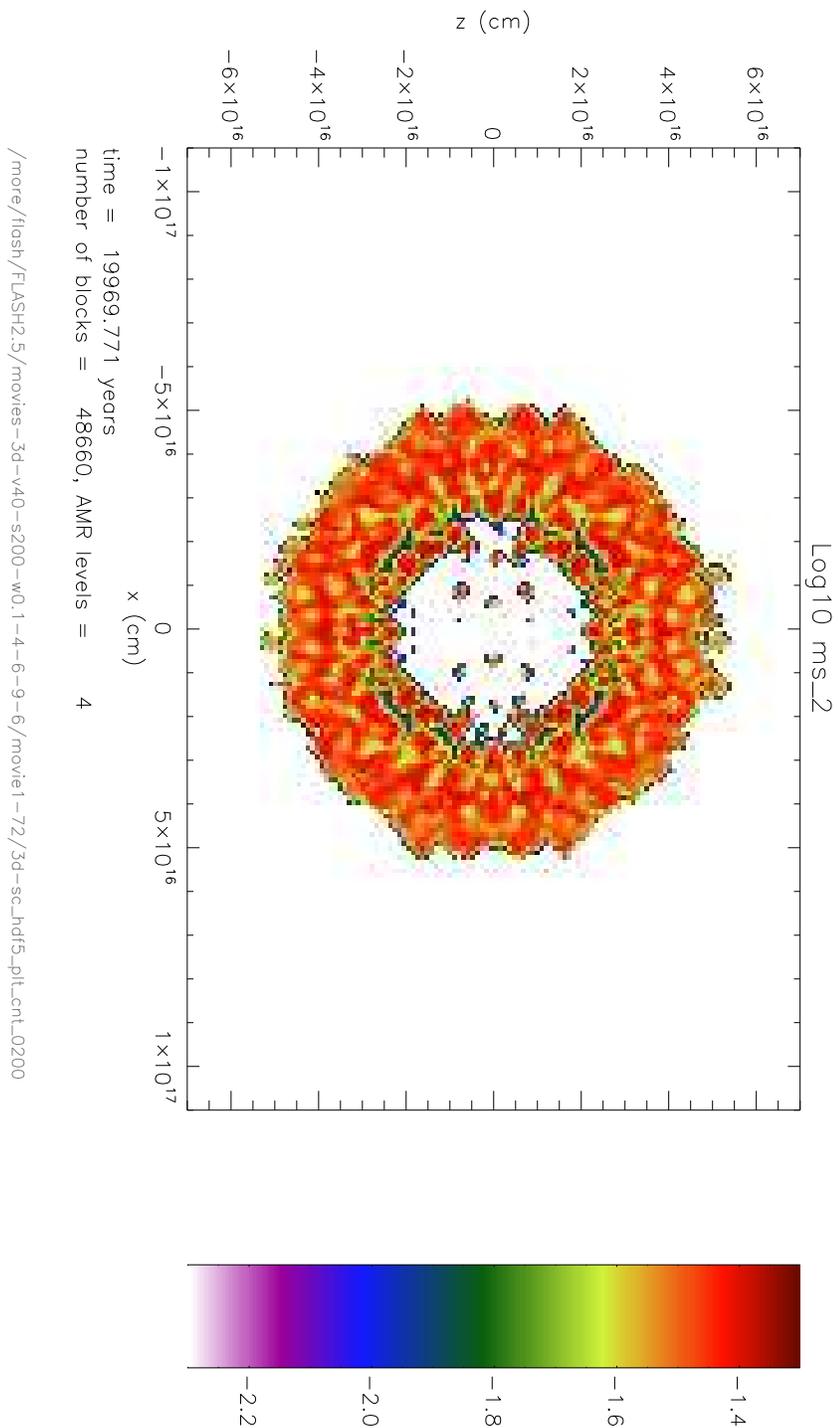}
\vspace{+0.25in}
\caption{Log10 of the color field distribution for model 40-200-0.1 
after $2.00 \times 10^{4}$ yrs of evolution, plotted in the $y = 0.11$
pc plane. The $x$ axis is horizontal and the $z$ axis is vertical.
Roughly 100 RT fingers are evident as the shock front material is
injected into the target cloud core.}
\end{figure}

\begin{figure}
\vspace{-1.0in}
\includegraphics[scale=.80]{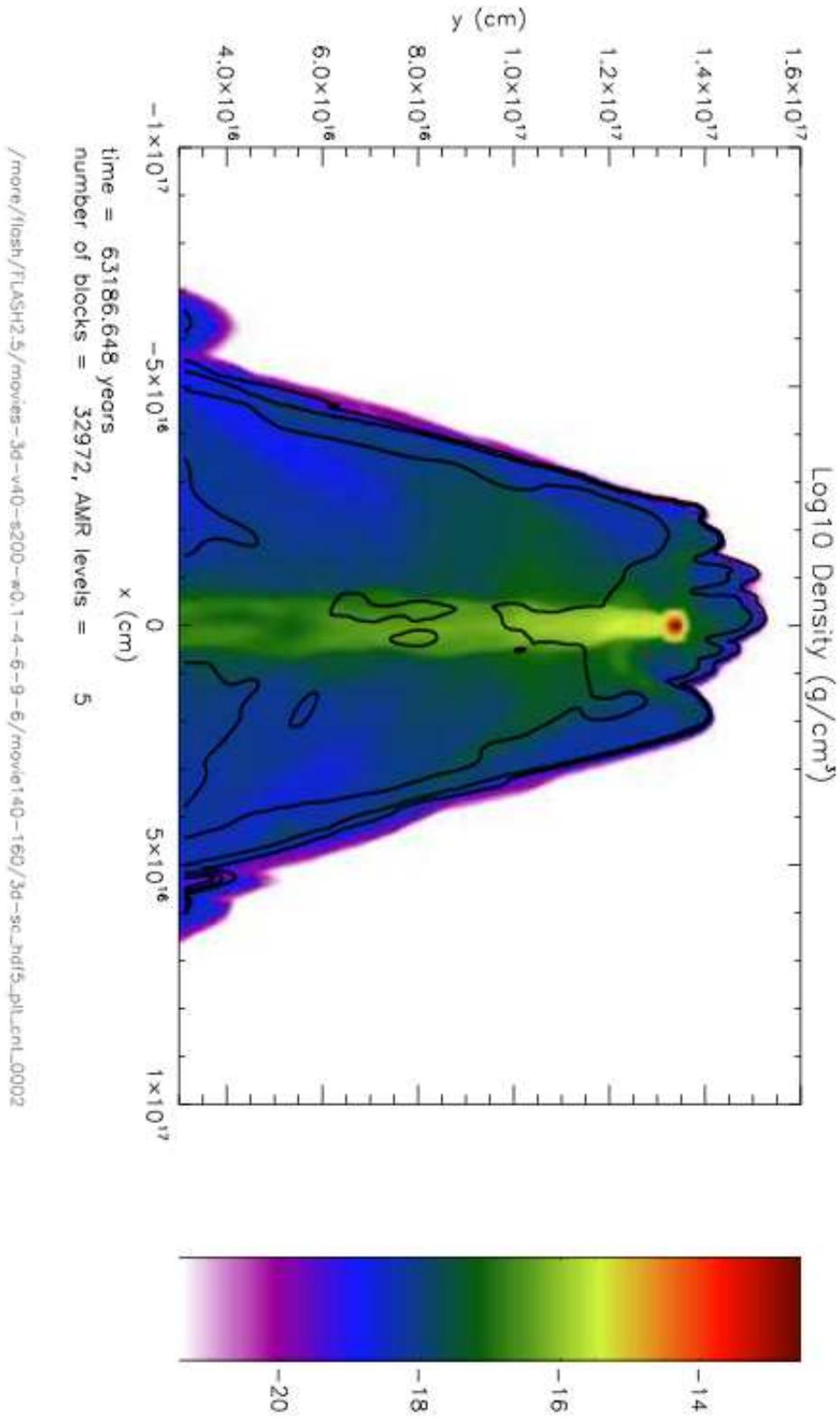}
\vspace{+0.25in}
\caption{Log10 of the density distribution and color field (SLRIs) 
contours (plotted in linear steps of 0.01) for model 40-200-0.1 
after $6.32 \times 10^{4}$ yrs of evolution, plotted as in 
Figure 1. The cloud core has been driven into dynamic collapse by the 
shock front, and has formed a well-defined, high density, collapsing 
protostar with a radius of $\sim$ 100 AU.}
\end{figure}

\begin{figure}
\vspace{-1.0in}
\includegraphics[scale=.80]{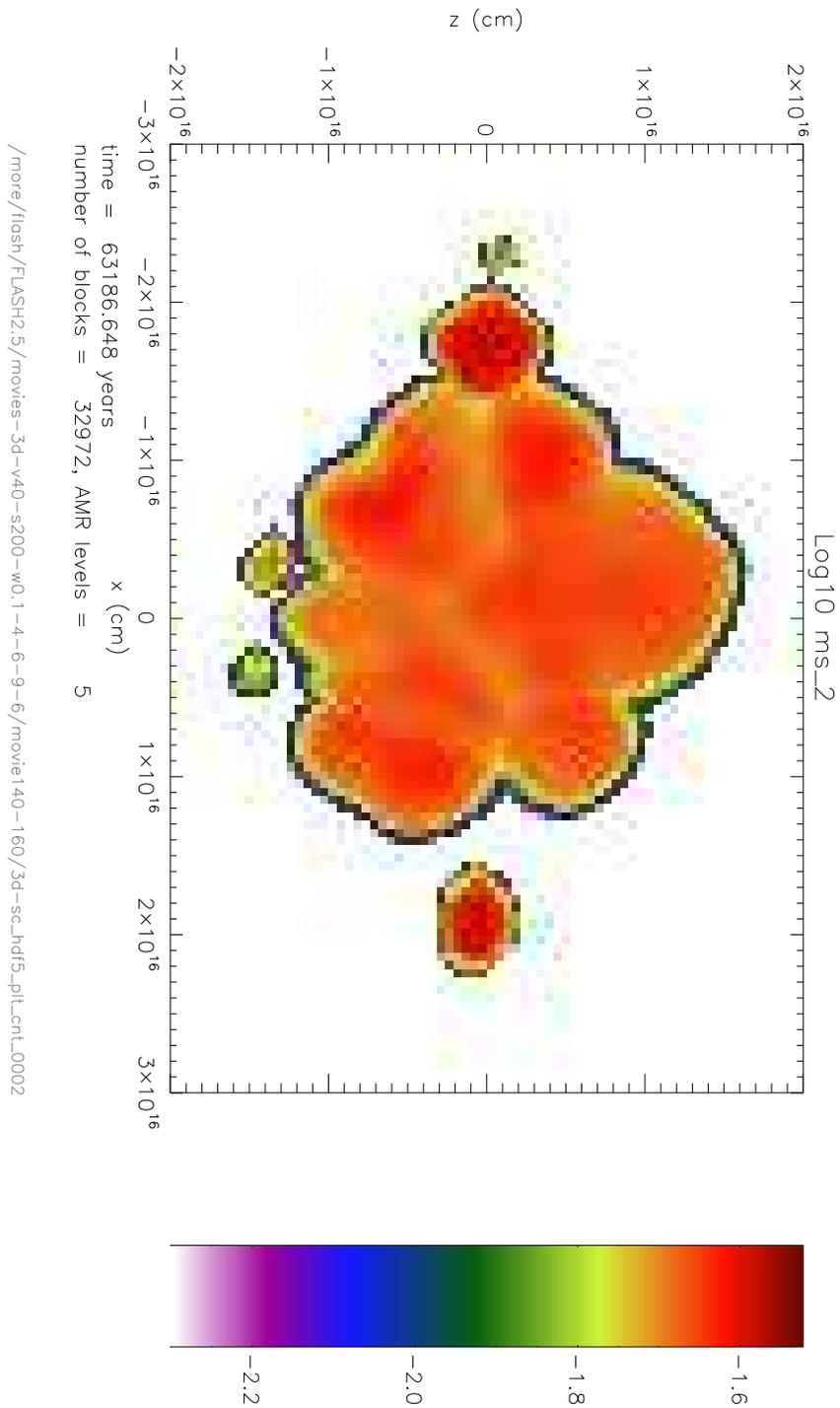}
\vspace{+0.25in}
\caption{Log10 of the color field distribution for model 40-200-0.1 
after $6.32 \times 10^{4}$ yrs of evolution, plotted as in Figure 2,
but in the $y = 0.045$ pc plane. Given the 2666 AU width of this plot,
and the $\sim 100$ AU size of the protostar (Figure 3) at this time, it 
is likely that only one (or at most a few) RT fingers will be responsible
for most of the shock wave injection into the collapsing presolar 
cloud core.}
\end{figure}

\end{document}